\documentclass[prl,aps,twocolumn,superscriptaddress]{revtex4-1}
\usepackage{amsfonts}

\usepackage{dcolumn}
\usepackage{bm}
\usepackage{tikz}
\usepackage[colorlinks=true,citecolor=blue,urlcolor=blue]{hyperref}
\usepackage{amsmath,amsfonts,amssymb,times,natbib}
\usepackage[standard]{ntheorem}

\begin{document}

\title{Experimental observation of dynamical bulk-surface correspondence for topological phases}

\author{Ya Wang}
\affiliation{Hefei National Laboratory for Physical Sciences at the Microscale and Department of Modern Physics, University of Science and Technology of China, Hefei 230026, China}
\affiliation{CAS Key Laboratory of Microscale Magnetic Resonance, University of Science and Technology of China}
\affiliation{Synergetic Innovation Center of Quantum Information and Quantum Physics, University of Science and Technology of China}
\author{Wentao Ji}
\affiliation{Hefei National Laboratory for Physical Sciences at the Microscale and Department of Modern Physics, University of Science and Technology of China, Hefei 230026, China}
\affiliation{CAS Key Laboratory of Microscale Magnetic Resonance, University of Science and Technology of China}
\affiliation{Synergetic Innovation Center of Quantum Information and Quantum Physics, University of Science and Technology of China}
\author{Zihua Chai}
\affiliation{Hefei National Laboratory for Physical Sciences at the Microscale and Department of Modern Physics, University of Science and Technology of China, Hefei 230026, China}
\affiliation{CAS Key Laboratory of Microscale Magnetic Resonance, University of Science and Technology of China}
\affiliation{Synergetic Innovation Center of Quantum Information and Quantum Physics, University of Science and Technology of China}
\author{Yuhang Guo}
\affiliation{Hefei National Laboratory for Physical Sciences at the Microscale and Department of Modern Physics, University of Science and Technology of China, Hefei 230026, China}
\affiliation{CAS Key Laboratory of Microscale Magnetic Resonance, University of Science and Technology of China}
\affiliation{Synergetic Innovation Center of Quantum Information and Quantum Physics, University of Science and Technology of China}
\author{Mengqi Wang}
\affiliation{Hefei National Laboratory for Physical Sciences at the Microscale and Department of Modern Physics, University of Science and Technology of China, Hefei 230026, China}
\affiliation{CAS Key Laboratory of Microscale Magnetic Resonance, University of Science and Technology of China}
\affiliation{Synergetic Innovation Center of Quantum Information and Quantum Physics, University of Science and Technology of China}
\author{Xiangyu Ye}
\affiliation{Hefei National Laboratory for Physical Sciences at the Microscale and Department of Modern Physics, University of Science and Technology of China, Hefei 230026, China}
\affiliation{CAS Key Laboratory of Microscale Magnetic Resonance, University of Science and Technology of China}
\affiliation{Synergetic Innovation Center of Quantum Information and Quantum Physics, University of Science and Technology of China}
\author{Pei Yu}
\affiliation{Hefei National Laboratory for Physical Sciences at the Microscale and Department of Modern Physics, University of Science and Technology of China, Hefei 230026, China}
\affiliation{CAS Key Laboratory of Microscale Magnetic Resonance, University of Science and Technology of China}
\affiliation{Synergetic Innovation Center of Quantum Information and Quantum Physics, University of Science and Technology of China}
\author{Long Zhang}
\affiliation{International Center for Quantum Materials, School of Physics, Peking University, Beijing 100871, China}
\affiliation{Collaborative Innovation Center of Quantum Matter, Beijing 100871, China}
\author{Xi Qin}
\affiliation{Hefei National Laboratory for Physical Sciences at the Microscale and Department of Modern Physics, University of Science and Technology of China, Hefei 230026, China}
\affiliation{CAS Key Laboratory of Microscale Magnetic Resonance, University of Science and Technology of China}
\affiliation{Synergetic Innovation Center of Quantum Information and Quantum Physics, University of Science and Technology of China}
\author{Pengfei Wang}
\affiliation{Hefei National Laboratory for Physical Sciences at the Microscale and Department of Modern Physics, University of Science and Technology of China, Hefei 230026, China}
\affiliation{CAS Key Laboratory of Microscale Magnetic Resonance, University of Science and Technology of China}
\affiliation{Synergetic Innovation Center of Quantum Information and Quantum Physics, University of Science and Technology of China}
\author{Fazhan Shi}
\affiliation{Hefei National Laboratory for Physical Sciences at the Microscale and Department of Modern Physics, University of Science and Technology of China, Hefei 230026, China}
\affiliation{CAS Key Laboratory of Microscale Magnetic Resonance, University of Science and Technology of China}
\affiliation{Synergetic Innovation Center of Quantum Information and Quantum Physics, University of Science and Technology of China}
\author{Xing Rong}
\affiliation{Hefei National Laboratory for Physical Sciences at the Microscale and Department of Modern Physics, University of Science and Technology of China, Hefei 230026, China}
\affiliation{CAS Key Laboratory of Microscale Magnetic Resonance, University of Science and Technology of China}
\affiliation{Synergetic Innovation Center of Quantum Information and Quantum Physics, University of Science and Technology of China}
\author{Dawei Lu}\email{ludw@sustech.edu.cn}
\affiliation{Shenzhen Institute for Quantum Science and Engineering and Department of Physics, Southern University of Science and Technology, Shenzhen 518055, China}
\affiliation{Center for Quantum Computing, Peng Cheng Laboratory, Shenzhen 518055, China}
\affiliation{Shenzhen Key Laboratory of Quantum Science and Engineering, Shenzhen 518055, China}
\author{Xiong-Jun Liu}\email{xiongjunliu@pku.edu.cn}
\affiliation{International Center for Quantum Materials, School of Physics, Peking University, Beijing 100871, China}
\affiliation{Collaborative Innovation Center of Quantum Matter, Beijing 100871, China}
\affiliation{Shenzhen Institute for Quantum Science and Engineering and Department of Physics, Southern University of Science and Technology, Shenzhen 518055, China}
\affiliation{Beijing Academy of Quantum Information Science, Beijing 100193, China}
\author{Jiangfeng Du}\email{ djf@ustc.edu.cn}
\affiliation{Hefei National Laboratory for Physical Sciences at the Microscale and Department of Modern Physics, University of Science and Technology of China, Hefei 230026, China}
\affiliation{CAS Key Laboratory of Microscale Magnetic Resonance, University of Science and Technology of China}
\affiliation{Synergetic Innovation Center of Quantum Information and Quantum Physics, University of Science and Technology of China}

\date{\today}
\begin{abstract}
We experimentally demonstrate a dynamical classification approach for investigation of topological quantum phases using a solid-state spin system through nitrogen-vacancy (NV) center in diamond. Similar to the bulk-boundary correspondence in real space at equilibrium, we observe a dynamical bulk-surface correspondence in the momentum space from a dynamical quench process. An emergent dynamical topological invariant is precisely measured in experiment by imaging the dynamical spin-textures on the recently defined band-inversion surfaces, with high topological numbers being implemented. Importantly, the dynamical classification approach is shown to be independent of quench ways and robust to the decoherence effects, offering a novel and practical strategy for dynamical topology characterization, especially for high dimensional gapped topological phases.
\end{abstract}


\maketitle
The topology of quantum systems has been developed into a major focus of research in physics since the discovery of quantum Hall effect \cite{Klitzing_Hall_1980, Tsui_Hall_1982}. Beyond the states of quantum matter characterized by Landau symmetry-breaking theory, the topological quantum phases bear a myriad of properties depending only on the topology \cite{Xiao_D_2010,Thouless_1982, Wen-XG_1990}, with the most celebrated paradigms discovered recently including the topological insulators~\cite{Kane_2005, Bernevig_2006,Konig_HgTe_2007, Hsieh_3D_2008, Xiao-Y_3D_2009, Chang_2013} and semimetals~\cite{Xu-SY_2015,Lv-BQ_2015}. Among the many exotic features emerging in a topological matter, the bulk-boundary correspondence is the most fundamental phenomenon showing that on the boundary gapless states can be obtained corresponding to and protected by the nontrivial topology in the bulk~\cite{Hasan_2010, Qi-XL_2011}.


At equilibrium, the number of topologically protected surface (edge) states is uniquely related to the bulk topological invariants~\cite{Hasan_2010, Qi-XL_2011}. This allows to detect topological insulators \cite{Konig_HgTe_2007, Hsieh_3D_2008, Xiao-Y_3D_2009, Chang_2013} and semimetals \cite{Xu-SY_2015,Lv-BQ_2015} by resolving the surface states from transport measurement or angle resolved photoemission spectroscopy. Apart from the direct measurement in condensed matter physics, quantum simulation may provide new strategies for the characterization \cite{Aidelsburger_2013,Liu-XJ_2013,  Miyake_2013, Atala_2013, Su-WP_1979, Jotzu_2014, Aidelsburger_2015, Lohse_2016, Nakajima_2016, Schweizer_2016, Song-B_2018,Flaschner_2016, Liu-XJ_Det_2013,  Wu-Z_2016}. For example, the band topology of 1D Su-Schrieffer-Heeger (SSH) chain can be determined by measuring the Zak phase\cite{Atala_2013,Su-WP_1979}, the bulk topology of a 2D Chern insulator can be observed by Hall transport studies \cite{Jotzu_2014, Aidelsburger_2015}, by Berry curvature mapping \cite{Flaschner_2016}, or by a minimal measurement strategy \cite{Liu-XJ_Det_2013} of imaging the spin texture at symmetric Bloch space\cite{Wu-Z_2016}. In addition, these systems naturally offer powerful probes and controllability in comparison with their condensed matter counterpart, which enables the study of non-equilibrium physics across topological phase transitions. A generic protocol in state-of-the-art experiments is to prepare a topologically trivial initial state and then observe the non-equilibrium dynamics after the Hamiltonian has been quenched into a topological regime. For such non-equilibrium process, the bulk-boundary correspondence, however, is no-longer valid \cite{DAlessio_2015, Caio_2015, Hu-Y_2016, Wilson_2016}. A natural challenge is to find a generic method to reveal dynamical signatures of topology. Recent works \cite{Heyl_2017, Tarnowski_2017, Wang_2017, ZHANG2018} start to study this question. In particular, a novel dynamical classification theory \cite{ZHANG2018}, which applies to a generic $d$-dimensional ($d$D) gapped topological phase, was proposed to address this fundamental issue theoretically. It shows that the equilibrium bulk topology universally corresponds to the dynamical topology emerging in the $(d-1)$D momentum subspace called band inversion surfaces (BISs). An initial experiment implementing this scheme has been performed in atomic systems \cite{Sun_2018} very recently. However, the essential dynamical bulk-surface correspondence, which necessitates to observe the dynamical topological invariant on BISs, is still not observed.

In this letter, we experimentally study the dynamical classification scheme with a diamond nitrogen-vacancy (NV) center spin system. We observe a dynamical bulk-surface correspondence~\cite{ZHANG2018} that the nontrivial bulk topology is related to a topological dynamical field emerging on BISs in the momentum subspace. Further, we show that this dynamical correspondence is not only universal in different quenching dynamics, but more importantly, invulnerable even in the presence of decoherence. Our results show a novel and practical strategy for dynamical topology characterization, which has broad applications in exploring topological quantum matter.

\begin{figure*}[t]
\centering
\includegraphics[width=1.8\columnwidth]{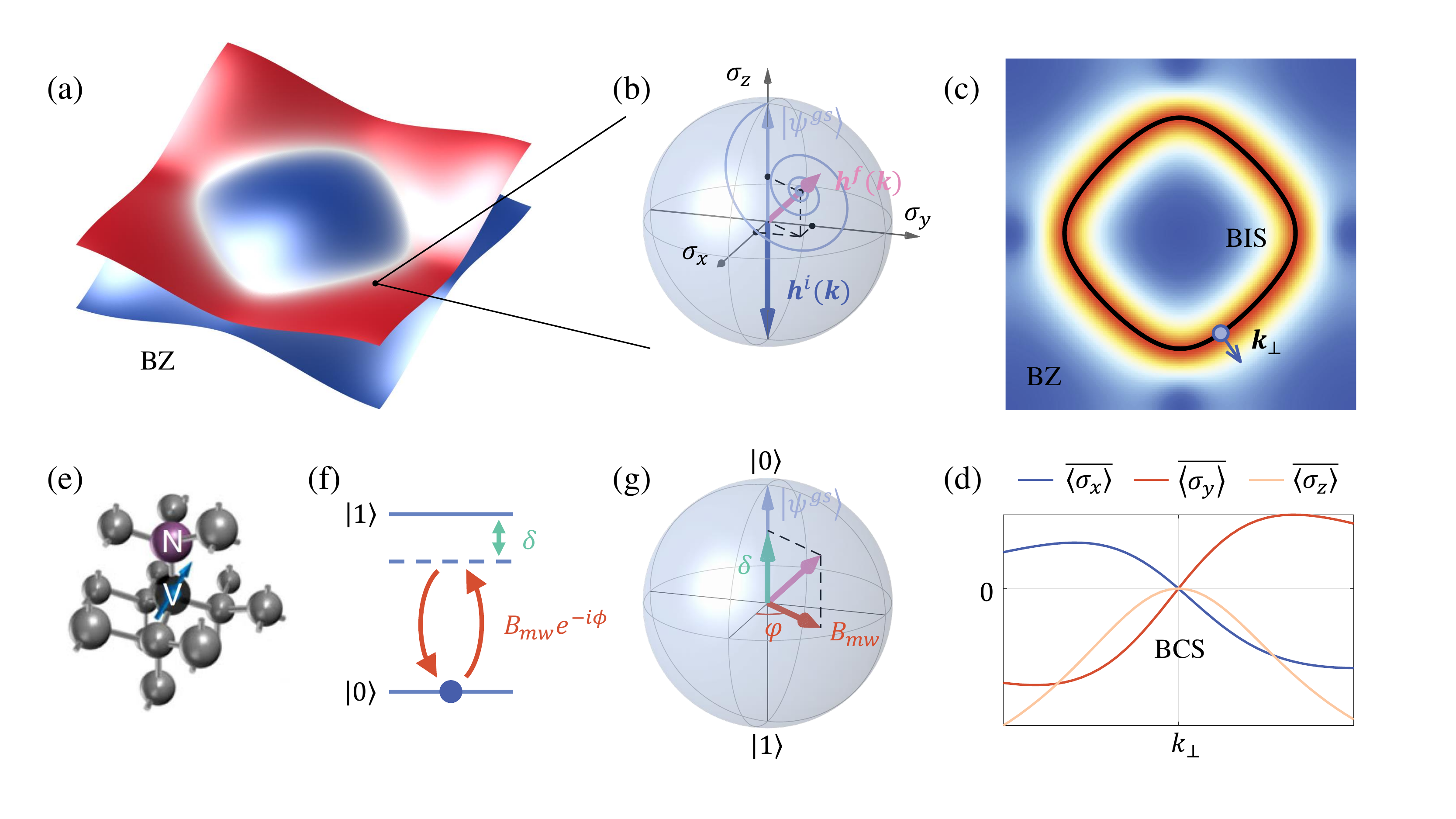}
\caption{Schematics of experimental dynamical classification protocol. (a) The band structure of a 2D quantum anomalous model in the topological regime. (b) Spin dynamics under quenching along the $z$ axis. The blue arrow denotes the initial ground state of the initial Hamiltonian $\mathcal{H}^{i}(\bold k)$. The pink arrow denotes the final Hamiltonian $\mathcal{H}^{f}(\bold k)$. The dynamics under other quenching directions can be visualized through a simple coordinate transformation. (c) The location of band inversion surface (BIS) detected through observing the time averaged spin polarization along the quenching direction (z axis here), if $\mathcal{H}^{f}$ is topological. $\textbf{k}_\bot$ denotes the norm factor of BIS. (d) Spin polarizations of different momentum points along $\textbf{k}_\bot$. (e) The experimental system: an NV center in the diamond. (f) The energy diagrams of the NV center used for simulation. A microwave pulse is applied to simulate the 2D quantum anomalous Hall model at each momentum via controlling its detuning frequency ($\delta$), amplitude ($B_{mw}$) and phase ($\phi$). (g) A visualization of the driven quantum system in the Bloch sphere.
}\label{scheme}
\end{figure*}

We consider the 2D quantum anomalous Hall model \cite{Liu-XJ_2014} realized in recent experiments \cite{Wu-Z_2016} to illustrate the scheme,
\begin{equation}\label{Hamiltonian}
\begin{aligned}
\mathcal{H_{QAH}} = (m_z & - 2t_0(\cos k_x + \cos k_y))\sigma_z \\
            &+ 2t_{so}(\sin k_x \sigma_x + \sin k_y \sigma_y),
\end{aligned}
\end{equation}
where $\textbf{k}=(k_x, k_y)$ represents the momentum, $t_0$ denotes the spin-conserved coefficients, and $t_{so}$ denotes spin-flip hopping coefficients corresponding to the 2D spin-orbit (SO) coupling, with $\sigma_{x,y,z}$ Pauli matrices. It is equivalent to an effective Zeeman field $\textbf{h(\textbf{k})}=(h_x, h_y, h_z)$ depending on the momentum $\textbf{k}$ applied in the Brillouin zone (BZ).
By tuning the term $m_z$, the quantum system is trivial for \textbf{$|m_z|>4t_0$} or topologically nontrivial for $|m_z| < 4 t_0$ and $m_z\neq0$. Fig.~\ref{scheme}(a) shows a typical band structure of the system in the topological regime, which exhibits band crossings. The general idea of our dynamical quenching method is summarized in Fig.~\ref{scheme}(b-d). Through a sudden change of $\textbf{h(\textbf{k})}$ from a trivial regime $\textbf{h}^i(\textbf{k})$ to a topological regime $\textbf{h}^f(\textbf{k})$, the initial quantum state (ground state of $\textbf{h}^i(\textbf{k})$) will evolve under the final Hamiltonian $\textbf{h}^f(\textbf{k})$). For example, Fig.~\ref{scheme}(b) displays a general momentum-dependent spin dynamics by quenching along the $z$ axis. The time averaged spin polarization (only $z$ component shows here) vanishes on a closed ring (Fig.~\ref{scheme}(c)), indicating the BIS for the 2D system, while near the BIS the dynamical spin polarizations ($x$ and $y$ components) can reconstruct the SO field and determine the topological invariant (Fig.~\ref{scheme}(d)).

We realize the scheme in a highly controllable solid-state system, which is a color defect named the NV center in diamond (Fig.~\ref{scheme}(e)) \cite{Marcus_2013}. The electrons around the defect form an effective electron spin with a spin triplet ground state (S = 1). In an external magnetic field along the NV axis, the electron spin states $m_s = \{0, +1, -1\}$ are well separated from each other and the states $m_s = 0, +1$ are used to study the two-band mode (Fig.~\ref{scheme}(f)). In the simulation, a green laser pulse is used to initialize the electron spin into $m_s = 0$ state which can then be prepared into the ground state of $\textbf{h}^i(\textbf{k})$ by unitary control, while a microwave pulse is applied to realize the final Hamiltonian. In the rotating frame, the Hamiltonian is written as $\mathcal{H}=\delta \sigma_z + B_{mw} (\cos \phi \sigma_x + \sin \phi \sigma_y)$, where $\delta$ is the detuning of the microwave frequency, $B_{mw}$ and $\phi$ are the driving amplitude and phase of the pulse. By choosing $\delta = m_z - 2t_0(\cos k_x + \cos k_y)$, $B_{mw} = 2 t_{so}$ and $\phi = \arctan (\sin k_y /\sin k_x)$, the electron spin of NV center in subspace spanned by ${m_s=0,+1}$ evolves exactly under the Hamiltonian $\mathcal{H_{QAH}}$ (Fig.~\ref{scheme}(g)).

The experiment was performed on a confocal setup at room temperature. The NV center was created by nitrogen ion implantation \cite{Meijer_2010} with an energy of 30 Kev. To improve the photon collection efficiency, a nanopillar structure \cite{Ali_2015, Fazhan_2018} was fabricated. An external static magnetic field around 370 Gauss was applied to remove the degeneracy between $m_s = +1$ and $m_s = -1$. Microwave pulses were irradiated via a coplanar waveguide and controlled by an arbitrary wave generator (CRS-AWG-C2S10N02, ChinaInstru $\&$ Quantum Tech). An arbitrary sequence generator (ASG-GT50-C, ChinaInstru $\&$ Quantum Tech) with 50 picosecond time resolution was used to control the timing sequences of the excitation laser and microwaves.

\begin{figure*}[t]
\centering
\includegraphics[width=1.8\columnwidth]{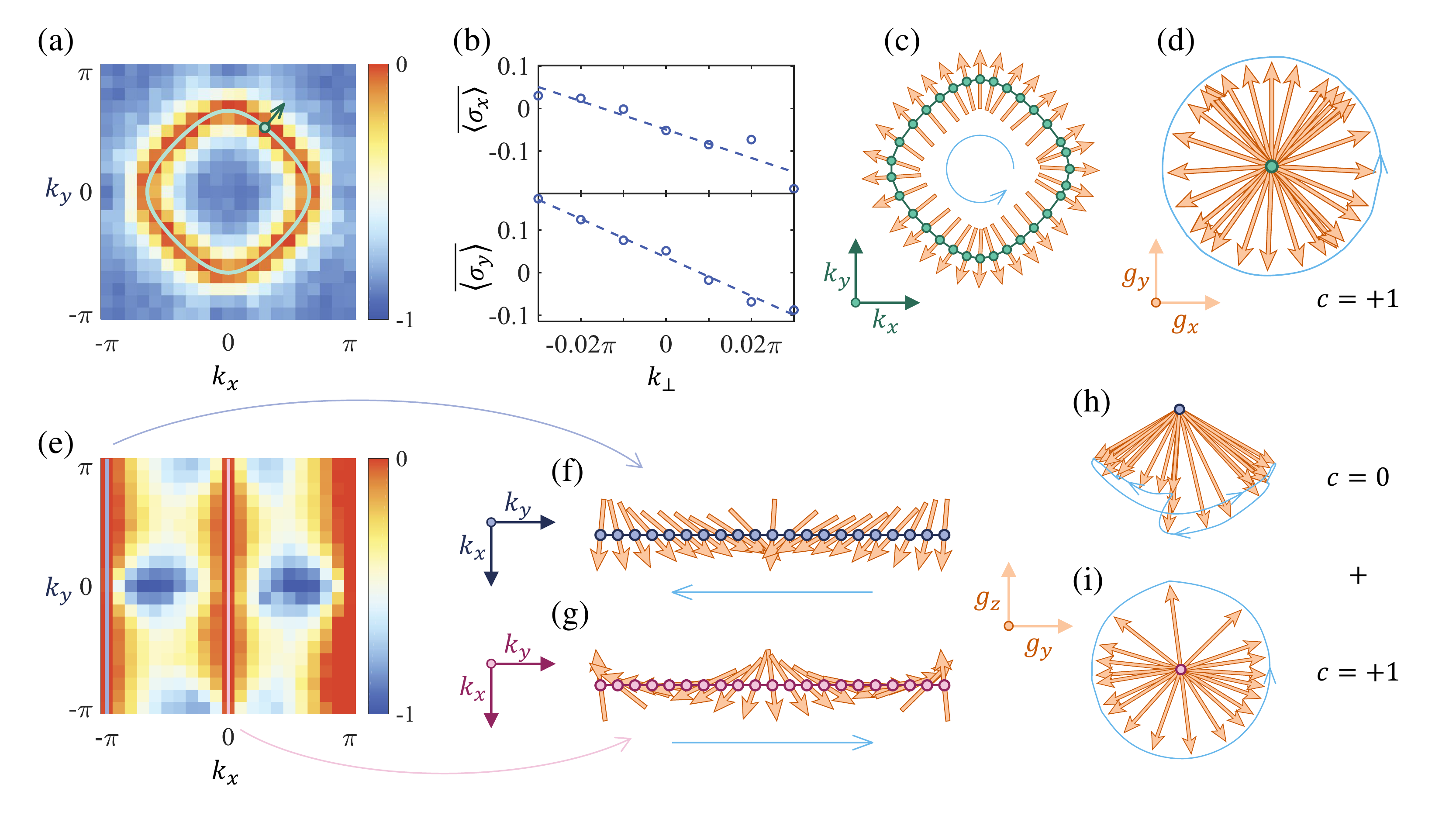}
\caption{(color online). Topological invariant emerging on BISs. (a) The BIS identified by measuring time-averaged spin polarization $\overline{\langle\sigma_z(\bold k,t)\rangle}$ in quenching along $z$ axis. (b) Linear dependence of spin polarizations $\overline{\langle\sigma_{x,y}\rangle}$ on momentum points along the $\vec{k}_{\bot}$ as shown in (a). The dashed lines are linear fit of the experimental results (dots). From the slop we obtain the corresponding $g_x$ and $g_y$. (c) The measured emergent dynamical field ($g_x,g_y,0$) on the BIS. (d) Redraw of (c) to count how many rounds the dynamical spin-texture field rotates around the central point. (e) The BIS of spin polarization $\overline{\langle\sigma_x\rangle}$ observed through quenching along the $x$ axis. (f,g) The measured spin-texture field ($g_0,g_y,g_z$) on the BIS. (h,i) The corresponding winding picture of (f) and (g). } \label{toponumber}
\end{figure*}

The quench study is performed by suddenly tuning the magnetization $m_z$ from an initial trivial phase into a topological regime. To measure the dynamical topological invariant emerging on BIS, we resolve the spin in a complete set of basis composed of $\sigma_{x,y,z}$ through a combination of microwave and laser pulses. The Rabi oscillations following quench are recorded at all $\bold k$ points, from which the time-averaged spin polarizations $\overline{\langle\sigma_{x,y,z}\rangle}$ are measured based on the first four periods.
In Fig.~\ref{toponumber} we consider two sets of examples for the detection. As shown in Fig.~\ref{toponumber}(a-d), we first quench the system along $z$ axis with the post-quench parameters being $m_z = t_0 = 2 t_{so}$. After determining the BIS by measuring $\overline{\langle\sigma_z\rangle}$, the time-averaged spin polarizations $\langle\sigma_{x,y}\rangle$ are measured at momentum points along the norm vector $\textbf{k}_\bot$ of BIS (Fig.\ref{scheme}(c)). Fig.~\ref{toponumber}(b) shows the linear area of Fig.~\ref{scheme}(b), from which one obtains a dynamical spin-texture field~\cite{ZHANG2018} $\vec{g}(\textbf{k})$ with the components $g_i(\textbf{k}) \equiv -\frac{1}{N_\textbf{k}}\partial_{k_\bot}\overline{\langle\sigma_i\rangle} (i= x, y)$ by measuring the slope, with $N_\textbf{k}$ the normalization factor. Fig.~\ref{toponumber}(c) shows the full imaging of this dynamical field, from which the topological invariant can be obtained through $\frac{1}{\pi}\int_{BIS} g_x dg_y = -\frac{1}{\pi}\int_{BIS}g_y dg_x$. Here instead we use the definition of winding number directly. Fig.~\ref{toponumber}(c) shows the visualization of the spin-texture field by redrawing them with the vector starting points jointed. One can count the number of revolutions to obtain the magnitude of topological invariant, and determine its sign according to the rotation direction. A counterclockwise full rotation is observed indicating that the winding number is $1$, directly corresponding to Chern number $c=1$ of the bulk.

Further, we implement the experimental measurement by quenching along $x$ axis~[Fig.~\ref{toponumber}(e-g)], with the parameters $m_z = 1.3 t_0 = 0.65 t_{so}$ being fixed. The Hamiltonian reads $\bold h(\bold k) = (m_x + 2t_{so} \sin k_x, 2t_{so} \sin k_y, m_z - 2t_0(\cos k_x + \cos k_y))$. The quenching parameter $m_x$ is suddenly varied from $m_x \gg t_0$ to 0. In this case the initial state is prepared as $(|m_s = 0\rangle + |m_s = +1\rangle)/\sqrt{2}$ through a green laser pulse followed by a $\pi_y/2$ pulse. The BISs (Fig.~\ref{toponumber}(e)) are measured by reading out the evolution in the $\sigma_x$ basis. This quenching leads to different configurations of BISs, which are identified as two lines at $k_x = -\pi$ and $k_x = 0$ connecting the edges of Brillouin zone~\cite{ZHANG2018}. The dynamical fields $\vec g(\bold k)$ on the BISs are further obtained from the measurements in the basis of $\sigma_z$ and $\sigma_x$. As shown in Fig.~\ref{toponumber}(f) and (g), the dynamical field shows a trivial pattern along $k_y$ at $k_x = -\pi$ but a nontrivial pattern at $k_x = 0$, giving the total winding number being $1$. These results show that while the configurations of BISs and quench dynamics are sharply different in different quench processes, the emergent topology in the quench dynamics is identical, showing from experiment that the dynamical characterization is universal.

\begin{figure}[t]
\centering
\includegraphics[width=1.0\columnwidth]{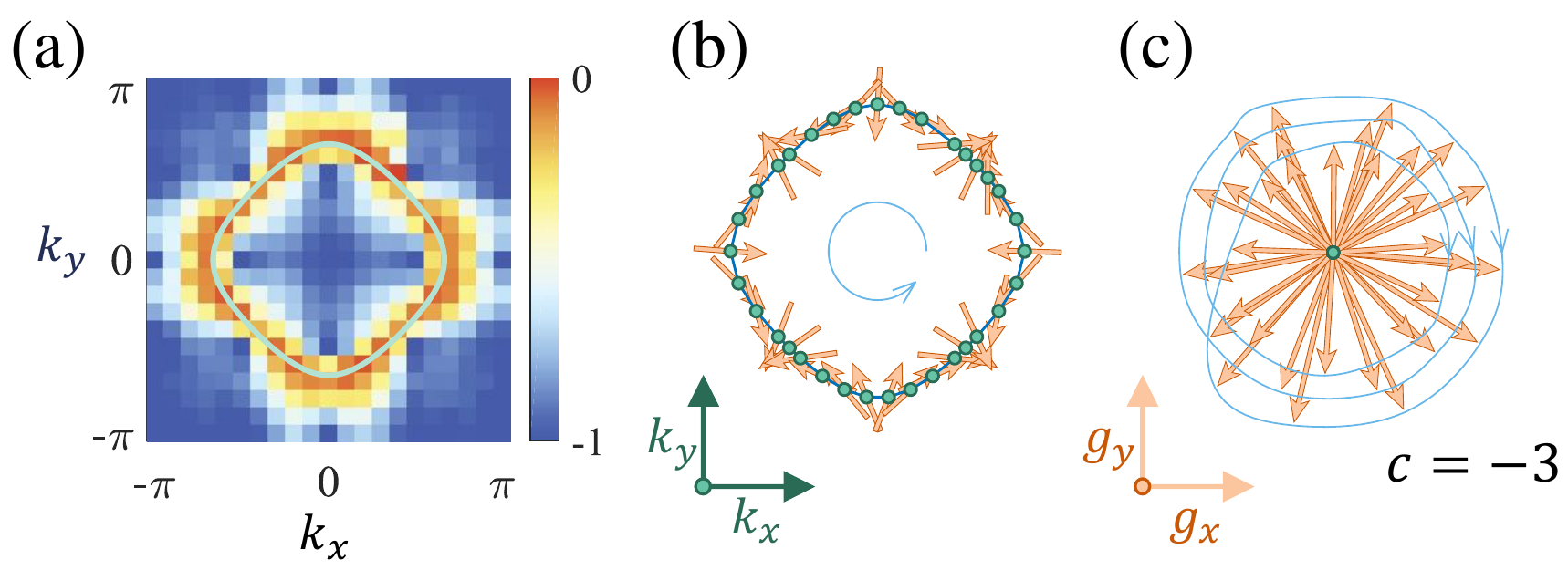}
\caption{(color online). Characterizing high topological invariant. (a) The BIS of spin polarizations $\overline{\langle\sigma_z\rangle}$ observed through quenching along the $z$ axis. (b) The measured spin-texture field ($g_x,g_y,0$) and (c) the corresponding winding number of (b).} \label{highnumber}
\end{figure}

We further apply this dynamical approach to characterize the topological phase with high Chern number $c=-3$. For this we implement the Hamiltonian that $\bold{h(\bold{k})} = (t_{so} \sin 2k_x, t_{so} \sin 2k_y, m_z - t_0(\cos k_x + \cos k_y))$, with $m_z\gg t_0$ before quench and $m_z = 0.5t_0 = t_{so}$ after quench. It can be shown that in equilibrium the lower Bloch band of the post-quench Hamiltonian has a high Chern number with $c=-3$. The measured results for quenching along $z$ axis are shown in Fig.~\ref{highnumber}. On BIS [see (a,b)], the length of the dynamical field $\vec{g}(\textbf{k})$ in Fig.~\ref{highnumber}(c) is varied for a clear display. Three full clockwise rotations along the closed BIS are explicitly observed, giving the dynamical winding number as $-3$, which corresponds to the 2D Chern number $c=-3$ of the bulk.

\begin{figure}[t]
\centering
\includegraphics[width=1\columnwidth]{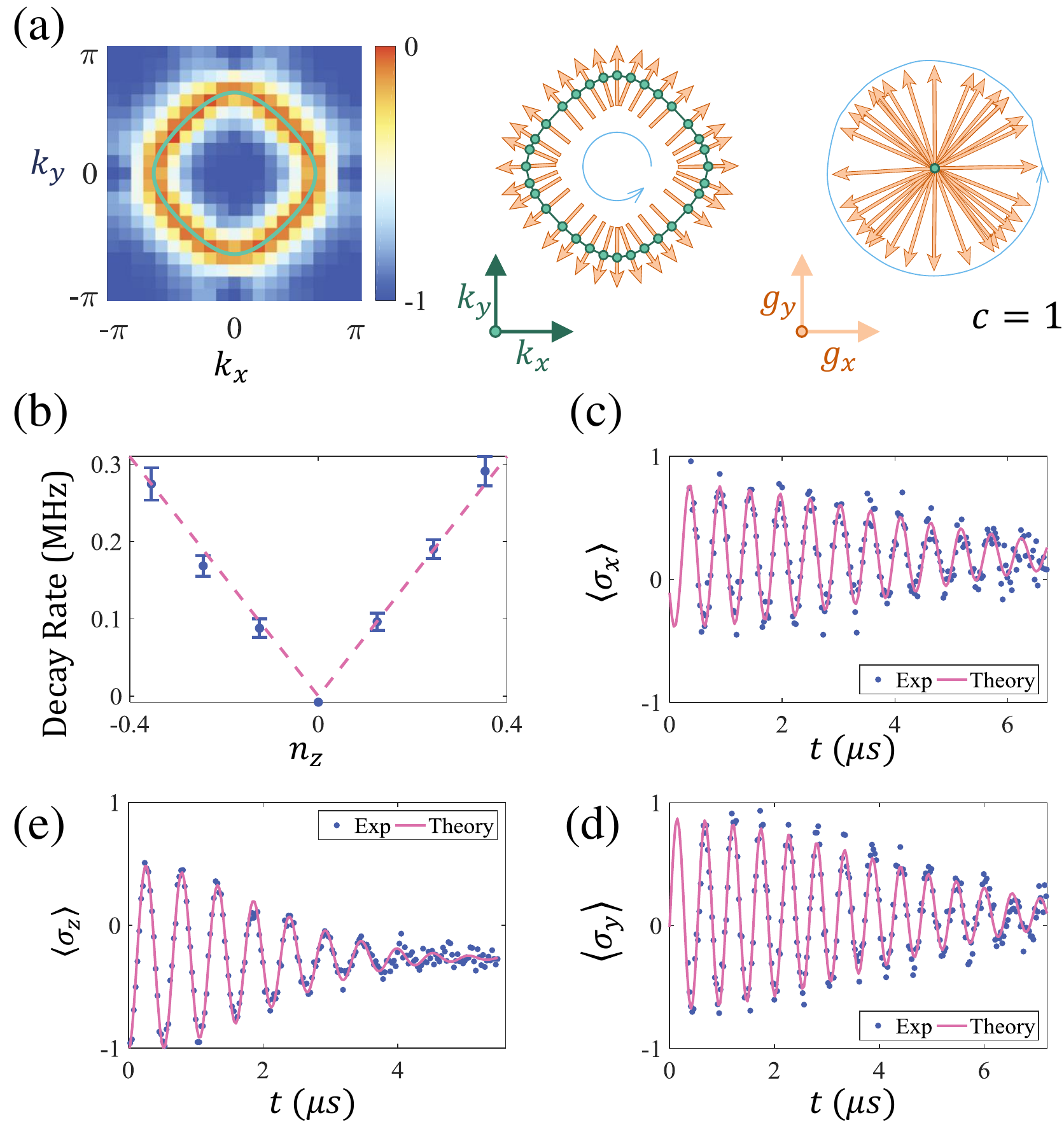}
\caption{(color online). Classifying topological invariant in the presence of decoherence. (a) The results of topological characterization in the presence of decoherence through quenching along the $z$ axis. The post-quench parameters are $m_z = t_0 = 2t_{so}$ as in (Fig.~\ref{toponumber}(a-d)) but with a smaller value of $t_{so} = 0.8$ MHz to enhance the effect of noise. Each pixel data is an average of the Rabi oscillation data more than 10 full periods as shown in (c). (b) Linear dependence of extracted decay rate on $n_z$ as observed by experiments and predicted by theory (Eq.~\ref{decoherence2}). (c,d,e) Rabi oscillations of $\langle\sigma_{x,y,z}\rangle$ which agree well with theoretical prediction, from which the decay rate is extracted.}\label{dephase}
\end{figure}

Finally, we explore an important issue, the validity of the dynamical classification in the presence of decoherence, which is a crucial issue generically existing in quantum dynamics. To investigate the effect of decoherence, we repeat the whole process as in Fig.~\ref{toponumber}(a-d), but with much longer evolution time. As shown in Fig.~\ref{dephase}(c-e), this induces an obvious decay in all spin components ($\sigma_{x,y,z}$) of the spin dynamics. Surprisingly, we find that the dynamical classification method is very robust to the decoherence. As shown in Fig.~\ref{dephase}(a), the BIS and topological invariant are almost the same as the results shown in Fig.~\ref{toponumber}(a-d). To gain a deep understanding of this excellent feature, we model the quench dynamics by considering the dominant dephasing noise arising from the nuclear spin bath around the central NV electron spin\cite{Du_2009,Zhao_2012}£º
\begin{equation}\label{decoherence1}
\begin{aligned}
\mathcal{H}_{dec} =  h_x \sigma_x + h_y \sigma_y + h_z \sigma_z + b_z \sigma_z.
\end{aligned}
\end{equation}
Here the last $b_z$-term in the right hand side represents the static magnetic noise which satisfies Gaussian distribution with the standard deviation $\sigma$. In the realistic case, we consider a relatively weak noise with strength $|b_z| \ll h=(h^2_x + h^2_y + h^2_z)^{1/2}$. The time-evolution of the spin polarization $\langle\sigma_i(\bold k)\rangle$ ($i=x,y,z$) following quench is given by
\begin{equation}\label{decoherence2}
\begin{aligned}
\langle\sigma_i(\bold k,t)\rangle= -h_ih_z/h^2+A_icos(ht+\varphi_i)e^{-\frac{\pi^2}{2}\sigma^2n_z^2t^2}
\end{aligned}
\end{equation}
where $n_z = h_z/h$ and $A_i, \varphi_i$ are determined by initial condition. As shown in Fig.~\ref{dephase}(b-c), this analytical form explains the experimental observations quite well. After time averaging, the time-dependent part vanishes and only the constant term is left, giving $\overline{\langle\sigma_{i}\rangle} = - h_ih_z/h^2$. Accordingly, the dynamical topology emerging on the BIS is not affected by the decoherence and well characterizes the post-quench topology.

The above result shows that the the present dynamical classification approach is fully immune to the dephasing effect. The essential reason is because the dynamical characterization adopts only the time-averaged terms, which naturally involve only the diagonal parts of the density matrix and insensitive to decoherence. Actually, the dephasing evolves an arbitrary initial state $|u(\bold k)\rangle=a(\bold k)|\uparrow\rangle+b(\bold k)|\downarrow\rangle$ to the mixed one described by a density matrix $\rho(t\to\infty)\rightarrow|a(\bold k)|^2|\uparrow\rangle\langle\uparrow|+|b(\bold k)|^2|\downarrow\rangle\langle\downarrow|$. With the mixed density matrix the spin polarizations at each momentum is unchanged, so is the topology. The high stability of the emergent dynamical topology on BIS may have potential applications in the future.

We have experimentally demonstrated a powerful dynamical classification approach in exploring topological quantum states. A dynamical bulk-surface correspondence was observed and demonstrated to be universal in different quenching dynamics. More importantly, we found that the emergent topology of quench dynamics is robust even in the presence of decoherence. As shown in a latest theory~\cite{LZhang2019}, the similar decoherence arises from correlation effects in the dynamical characterization of an interacting topological system. Our study implies that the present dynamical simulator based on NV center may be useful to further emulate effectively the correlation effects in the topological characterization. This study can be generalized to high dimensional gapped topological phases.

\emph{Acknowledgement.-}
This work is supported by the National Key R$\&$D Program of China (Grant No. 2018YFA0306600, 2017YFA0305000, 2016YFA0301604, 2016YFB0501603), the NNSFC (Grants No. 11775209, 81788101, 11761131011, 11761161003, 11722544, 11574008, 11825401, 11605005, 11875159, U1801661), the CAS (Grants No. GJJSTD20170001, No. QYZDY-SSW-SLH004, No. QYZDB-SSW-SLH005), Anhui Initiative in Quantum Information Technologies (Grant No. AHY050000), the Fundamental Research Funds for the Central Universities, the Innovative Program of Development Foundation of Hefei Center for Physical Science and Technology (Grants No. 2017FXCX005), Science, Technology and Innovation Commission of Shenzhen Municipality (Grants No. ZDSYS20170303165926217, JCYJ20170412152620376), Guangdong Innovative and Entrepreneurial Research Team Program (Grant No. 2016ZT06D348), the Thousand-Young-Talent Program of China and the Youth Innovation Promotion Association of Chinese Academy of Sciences.

\bibliography{references}

\end{document}